\documentclass[a4paper,twocolumn,11pt,published,accepted=2021-03-03]{quantumarticle}
\pdfoutput=1
\usepackage[utf8]{inputenc}
\usepackage[english]{babel}
\usepackage[T1]{fontenc}
\usepackage{amsmath}
\usepackage{hyperref}
\usepackage{color}
\usepackage{graphicx}
\usepackage{dcolumn}
\usepackage{bm}
\usepackage{physics}
\usepackage{hyperref}
\usepackage{url}
\usepackage{subcaption}
\usepackage{amsmath}
\usepackage{ dsfont }
\usepackage{verbatim}
\usepackage{amssymb}
\usepackage{amsfonts}              
\usepackage{amsthm}                
\usepackage{mathtools}
\usepackage{kotex}
\usepackage{qcircuit}
\usepackage[normalem]{ulem}
\usepackage[numbers,sort&compress]{natbib}

\newtheorem{thm}{Theorem}
\newtheorem{lem}[thm]{Lemma}

\newtheorem{theorem}{Theorem}

\newtheorem{definition}[theorem]{Definition}

\newcommand{\remark}[1]{{#1}}

\usepackage{tikz}
\usepackage{lipsum}
\begin{document}

\title{\remark{Quantum one-time tables} for unconditionally secure qubit-commitment}

\author{Seok Hyung Lie}
\affiliation{%
 Department of Physics and Astronomy, Seoul National University, Seoul, 151-742, Korea
}%
\author{Hyukjoon Kwon}
\affiliation{QOLS, Blackett Laboratory, Imperial College London, London SW7 2AZ, United Kingdom}
\author{M. S. Kim}
\affiliation{QOLS, Blackett Laboratory, Imperial College London, London SW7 2AZ, United Kingdom}
\affiliation{Korea Institute for Advanced Study, Seoul, 02455, Korea}
\author{Hyunseok Jeong}\email{h.jeong37@gmail.com}
\affiliation{%
 Department of Physics and Astronomy, Seoul National University, Seoul, 151-742, Korea
}%

\date{2020-08-31}

\begin{abstract}

The commodity-based cryptography is an alternative approach to realize conventionally impossible cryptographic primitives such as unconditionally secure bit-commitment by consuming pre-established correlation between distrustful participants. A unit of such classical correlation is known as the one-time table (OTT). In this paper, we introduce a new example besides quantum key distribution in which quantum correlation is useful for cryptography. We propose a scheme for unconditionally secure qubit-commitment, a quantum cryptographic primitive forbidden by the recently proven no-masking theorem in the standard model,
based on the consumption of the quantum generalization of the OTT, the bipartite quantum state we named \textit{quantum one-time tables} (QOTT). The construction of the QOTT is based on the newly analyzed internal structure of quantum masker and the quantum secret sharing schemes. Our qubit-commitment scheme is shown to be universally composable. We propose to measure the randomness cost of preparing a (Q)OTT in terms of its entropy, and show that the QOTT with superdense coding can increase the security level with half the cost of OTTs for unconditionally secure bit-commitment. The QOTT exemplifies an operational setting where neither maximally classically correlated state nor maximally entangled state, but rather a well-structured partially entangled mixed state is more valuable resource.
\end{abstract}

\maketitle

\section{Introduction}

In a commitment protocol, Alice commits to a secret value by transmitting an encoding of the value to Bob.
If Bob cannot access the value until revealed by Alice, the scheme is said to be secure against Bob. On the other hand, if Bob can reject Alice's cheating of revealing a value different from the originally committed value, then the scheme is said to be secure against Alice. An unconditionally and perfectly secure \cite{Rivest} commitment scheme could have many cryptographical applications \cite{Rabin, Goldreich}. However, such a commitment protocol is impossible since the perfect securities against Alice and Bob are incompatible. Quantum bit-commitment is an attempt to circumvent this difficulty by using quantum mechanics  \cite{Brassard}. However, it was proved \cite{M,LC} that an unconditionally secure commitment of a classical value is impossible even with the aid of quantum mechanics \remark{unless there is a relativistic structure that imposes causal restrictions between prover and verifier \cite{kent1999unconditionally,lunghi2013experimental,kent2012unconditionally}}.

To circumvent this difficulty, a new approach called the \textit{commodity-based cryptography} \cite{beaver1997commodity} was developed. Since secure two-party computation is impossible without mutual trust, the suggested idea was to construct cryptographical primitives that consume tradeable resource named the \textit{one-time table} (OTT). The OTT is a unit of suitably pre-calculated correlation that provides verifiable randomness to mutually mistrustful clients. The OTTs enable unconditionally secure bit-commitment, oblivious transfer \cite{Rivest} and field computation \cite{beaver1998one}. While the OTT could be established off-line (before a useful primitive protocol begins) by a central server (known as the `trusted initializer' \cite{Rivest}), also there has been a recent attempt to construct a protocol for establishing the OTT without a third party \cite{yu2019quantum}. Therefore, we can treat the OTTs as a resource stored in the form of correlation regardless of its origin.
 
The OTTs studied so far are, however, all classical correlations. This raises a natural question that if there is a quantum generalization of the OTT that is more suitable for quantum two-party tasks. In this paper, we construct  the \textit{quantum one-time table} (QOTT) for a universally composable qubit-commitment Qubit-commitment is a quantum cryptographic primitive that is impossible utilizing only pure states because of the recently proven no-go result known as the no-masking theorem \cite{KM}. The construction of the QOTT is based on the internal structure of quantum masker reinterpreted as a quantum process that consumes randomness to hide quantum information into bipartite correlations, and its relation with quantum secret sharing protocols \cite{cleve1999share, gottesman2000theory}. From this, we show that the no-masking theorem cannot be extended to mixed states in the way that the no-cloning theorem \cite{WZ} was extended to the no-broadcasting theorem \cite{BCFHS}, and that the qualitatively stronger constraint on the strength of viable quantum correlation requires more randomness for masking quantum information.

\remark{We suggest the entropy of commodity such as (Q)OTT, named the \textit{shared randomness cost}, as a measure of the randomness cost of a commodity-based cryptography protocol. We show that our QOTT-based qubit-commitment, which is different from quantum bit-commitment as will be elaborated afterwards, scheme could achieve asymptotically the same shared randomness cost compared to Rivest's OTT-based bit-commitment scheme \cite{Rivest}. When the superdense coding is employed, this implies that the QOTT-based bit-commitment scheme has half the randomness cost of the OTT-based bit-commitment scheme.}

 \begin{figure*}[t]
\captionsetup[subfigure]{justification=centering}
\begin{subfigure}{.25\textwidth}
\centering
$$\Qcircuit @C=.5em @R=.5em{
  \lstick{} & \lstick{\rho_{\text{in}}}& \gate{X} & \ctrl{2} & \qw &\rstick{A_1} \qw    \\
  & \lstick{\ket{0}}& \qw & \qw & \gate{X} &\rstick{A_2} \qw    \\
  & \lstick{\ket{0}}& \qw & \gate{X} & \qw &\rstick{B_1} \qw    \\
  &  \lstick{\frac{1}{d}\mathds{1}}& \ctrl{-3} & \gate{H} & \ctrl{-2} &\rstick{B_2} \qw  
}$$\centering \caption{The 4-qubit(qudit) masker.}
\end{subfigure}\begin{subfigure}{.25\textwidth}
\centering
$$\Qcircuit @C=1.2em @R=2.2em{
    \lstick{} & \lstick{\rho_{\text{in}}}& \gate{X} & \gate{Z} &  \rstick{A_1} \qw  \\
    &\lstick{\frac{1}{d}\mathds{1}}& \ctrl{-1} & \qw &  \rstick{B_1} \qw \\
    &\lstick{\frac{1}{d}\mathds{1}}& \qw & \ctrl{-2} & \rstick{B_2} \qw
}$$ \caption{The quantum one time pad (QOTP).}
\end{subfigure}\begin{subfigure}{.25\textwidth}
$$\Qcircuit @C=.6em @R=3em{
    \lstick{} & \lstick{\rho_{\text{in}}}& \gate{X}  & \qw & \ctrl{1} & \rstick{A} \qw  \\
    &\lstick{\frac{1}{d}\mathds{1}}& \ctrl{-1}  & \qw & \gate{X}& \rstick{B} \qw 
}$$\caption{A minimal masker.}
\centering
\end{subfigure}\begin{subfigure}{.25\textwidth}
$$\Qcircuit @C=.2em @R=3em{
    \lstick{} & \lstick{\rho_{\text{in}}}& \gate{H^2} &\gate{X^2} & \ctrl{1}  & \qw  & \rstick{B} \qw  \\
    &\lstick{\frac{1}{d}\mathds{1}}& \qw &\ctrl{-1} & \gate{X^\dag} & \gate{H^2} & \rstick{K} \qw 
}$$\caption{A minimal masker that is dual to (c) for partitions $(A,B)$ and $(B,K)$.}
\centering
\end{subfigure}

    \caption{Examples of universal quantum maskers. Output systems with indices $A_i$ and $B_i$ belong to Alice and Bob, respectively. In each case, discarding one party's quantum state yields the maximally mixed state for the other party. $\rho_{\text{in}}$ denotes the input state and $\frac{\mathds{1}}{d}$ denotes the maximally mixed state. \remark{(a)} The output state is partially entangled in general. \remark{(b)}  The output state is always a classical-quantum state therefore separable. \remark{(c) A minimal masker for any odd $d$, where every system has the minimal dimension $d$ and has a maximally mixed marginal state. By purifying the safe system $S$ of (c) to $SK$ and tracing out system $A$, one gets (d). We will call such maskers are dual to each other for given partitions. Here, $X:=\sum_{j=1}^d \dyad{j\oplus 1 \; (\text{mod } d)}{j}$, $Z:=\sum_{j=1}^d \exp({i2\pi j}/{d})\dyad{j}$ and $H:=\frac{1}{\sqrt{d}}\sum_{j,k} e^{i2\pi jk/d} |j\rangle \langle k|$ and the controlled-$G$ gate for a set of operators $\{G^j\}$ is defined as $\sum_j \dyad{j}\otimes G^j$.}  }
    \label{fig:1}
\end{figure*}

\section{Quantum Masker} 
When it comes to commitment schemes, security against Bob, also known as the \textit{hiding property}, is important. However, for qubit-commitment schemes, the information being committed to should be also hidden from Alice, otherwise Alice could freely change the information \cite{gottesman2000theory}. It is because, from the no-cloning theorem, acquisition of quantum information implies being only one who is holding that information, therefore modification of that information cannot be detected. Therefore it is natural to design a qubit-commitment scheme based on \textit{quantum maskers.}

\textit{Masking quantum information} is a quantum process that encodes a quantum state in a bipartite quantum system, while hiding it from both subsystems. Quantum masker was first introduced \cite{KM} for  pure bipartite states, where entanglement is the only form of correlations. However, there are \textit{quantum correlations beyond entanglement} \cite{QC,QCBE,diss} in the case of mixed states. A typical example  is quantum discord \cite{Dis}. We redefine the quantum masker for a general mixed state. Let $\mathcal{B(H)}$ be the algebra of bounded operators on a Hilbert space $\mathcal{H}$.

\begin{definition}  An operator $\mathcal{M}$ from $\mathcal{B}(\mathcal{H}_A)$ to $\mathcal{B}(\mathcal{H}_{A'}\otimes\mathcal{H}_{B'})$ is said to mask quantum information contained in states $\{{\phi_k}_A\in\mathcal{B}(\mathcal{H}_A)\}$ by mapping them to $\{{\Psi_k}_{A'B'}\in\mathcal{B}(\mathcal{H}_{A'}\otimes\mathcal{H}_{B'})\}$ such that all marginal states of ${\Psi_k}_{A'B'}=\mathcal{M}(\phi_{kA})$ are identical, i.e., $\rho_{A'}=\Tr_{B'}{{\Psi_k}_{A'B'}},$
and $\rho_{B'}=\Tr_{A'}{{\Psi_k}_{A'B'}}$,  with an unmasking operator $\mathcal{U}$ from $\mathcal{B}(\mathcal{H}_{A'}\otimes\mathcal{H}_{B'})$ to $\mathcal{B}(\mathcal{H}_A)$ such that
$$\mathcal{U}({\Psi_k}_{A'B'}) = {\phi_k}_A$$
for all $k$. 
\end{definition}
\remark{
 We call such $\mathcal{M}$ a (quantum) masker, and we say that $\mathcal{M}$ is universal if it masks an arbitrary quantum state. Our main interest is universal quantum masker, so unless remarked otherwise, afterwards every masker is assumed to be universal.
 
 Here, two observations can be made.} First, every universal quantum masker is demanded to be an invertible quantum process. Such a process, say $\Phi_M$, can always be expressed \cite{NS} in terms of a quantum state $\omega_S$ and a unitary transformation $M$ which maps the input system $C$ and the ancillary system $S$ to the systems $A$ and $B$ such that
 \begin{equation} \label{eqn:1}
     \Phi_M(\rho_C)=M_{CS\to AB}(\rho_C \otimes \omega_S)M_{AB \to CS}^\dag,
 \end{equation}
 for every quantum state $\rho_C$ with some ancillary state $\omega_S$. In the equations given hereunder, system subscripts will be omitted when it is clear from the context. We denote the unitary $M$ as the masking unitary of $\Phi_M$ and the ancillary state $\omega_S$ as the \textit{safe} state of $\Phi_M$. \remark{This observation implies that masking quantum information is a process that should consume randomness supplied in the form of safe state. This raises natural questions : Is it really possible to mask quantum information when randomness is supplied? If so, how much is needed? Asking this question is appropriate given the demand in computer science and quantum information science for adopting the perspective treating randomness as a resource \cite{Boykin, Brian, Boes, NS}.

 The second observation answers these questions. The definition of masking quantum information is equivalent to $(2,2)-$threshold quantum secret sharing (QSS) scheme \cite{cleve1999share, gottesman2000theory}, thus possible. (See Fig. \ref{fig:1} for examples.) Also, we can see that every purification of a quantum masker is a $(2,3)-$threshold QSS scheme, as a consequence of the no-hiding theorem \cite{braunstein2007quantum}. From the expression (\ref{eqn:1}), a purification of a masking process $\Phi_M$ can be acquired by purifying its safe state $\omega_S$ into a purification $\ket{\Omega}_{SK},$ with a purifying system $K$. We will call such a system as the \textit{key} system of the quantum masker and the state $\ket{\Omega}_{SK}$ as the \textit{safe-key} state.
 
Before constructing a qubit-commitment scheme based on quantum masking unitaries, we finish this section by introducing some analysis of the randomness cost of quantum maskers. Readers can skip Theorem \ref{thm2} and still understand the next section. Since any unauthorized subsystem of any $(k,n)-$threshold QSS scheme hiding $d-$dimensional quantum state should be in a constant quantum state (regardless of the secret state) with von Neumann entropy no less than $\log_2d$ bits, \cite{imai2003quantum}, so the key system of a quantum masker should, too. Since the safe-key state is a pure entangled state, this gives the lower bound of the randomness cost, i.e. the von Neumann entropy of the safe state. In Theorem \ref{thm2}, the minimal randomness costs of masking quantum information into various types of quantum correlations are given. Detailed proofs of all the theorems throughout this paper are provided in Appendix.
 
 }
\begin{theorem} \label{thm2}
Let $\omega_S$ be the safe state of a universal quantum masker $\Phi_M$ for a $d$-dimensional quantum system. 
The von Neumann entropy of $\omega_S$ is (i) no less than $\log_2d$ bits  if there is no other constraint, (ii) strictly larger than $\log_2d$ bits when any output state of $\Phi_M$ should be separable, and (iii) no less than $2\log_2 d$ bits when any output state of $\Phi_M$ should be quantum-classical. (iv) There is no such $\Phi_M$ that any output state of $\Phi_M$ is classically correlated.
\end{theorem}

The part \textit{(i)} yields the no-masking theorem as its corollary, as there are no universal quantum maskers that consume zero amount of randomness. \textit{(ii)} and \textit{(iii)} show that with the more restricted the kind of correlation, the more randomness is required to mask a quantum state, while \textit{(iv)} implies that quantum correlation is indispensable for masking quantum information. The last result \textit{(iv)} has an interesting implication for QSS that it is impossible to share a quantum secret among three parties exploiting only genuine tripartite entanglement. It is because a purification of classically correlated bipartite quantum state is always a genuine tripartite entangled state and the converse is also true \cite{hayashi2011weaker}.

\section{Quantum One-Time Tables for Qubit-Commitment} \label{sec:3}
 Based on the properties of the quantum masker, we construct a new unit of quantum commodity that can be used for an unconditionally secure qubit-commitment scheme. When commitment schemes are concerned, \remark{we will call a scheme bit-commitment scheme if the secret value is classical information, (`bit string'), and if the secret value is a quantum state, we will call it similarly \textit{qubit-commitment scheme}, even if the state is not 2-dimensional. Note that it is different from quantum bit-commitment scheme, which is a special case of bit-commitment scheme realized with quantum mechanical method. Also, we will call a commitment scheme $d-$dimensional if the secret value is $d-$dimensional.}
 
 Let us briefly review unconditionally secure classical bit-commitment using the OTT by Rivest . Let $p$ be an arbitrary prime number. The OTT used in Ref. \cite{Rivest} for $p-$dimensional bit-commitment scheme is a classical bipartite state shared between two parties composed of two parts: \textit{(i)} two random numbers in $\mathds{Z}_p$, $(a,b)$, generated from each party and their copies  respectively kept by Alice and Bob, and \textit{(ii)} a function of of the two random numbers $(a,b)$ (e.g. product modulo $p$) hidden and shared between Alice and Bob through the one-time pad (OTP) cipher, e.g. $(r,r+a\cdot b)$ with a random number $r \in \mathds{Z}_p$ \cite{Rivest}. The piece of information $a$ and $b$ held by  respectively Alice and Bob can be interpreted as the reference systems for their own random numbers i.e. maximally correlated with their counterparts.
 
\begin{figure}
    
    \includegraphics[width=.5\textwidth]{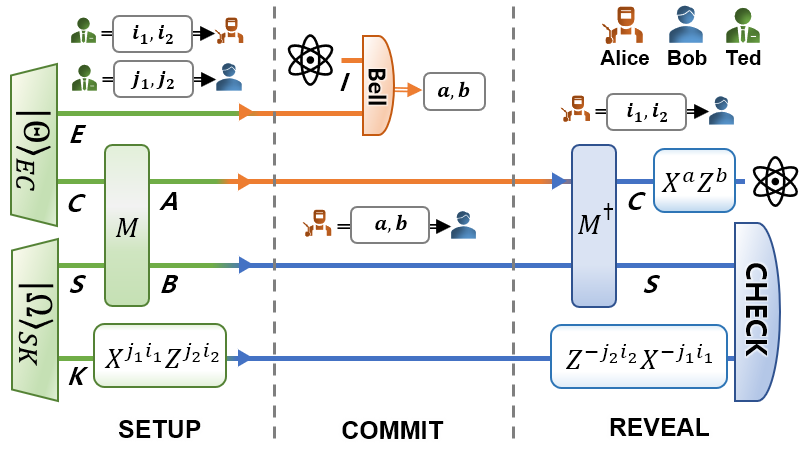}
    \centering
    \caption{Diagrammatic description of the initializer scheme. The color of each system represents its possessor (see the top-right corner) at each phase of the scheme and the arrows denote transmissions of quantum systems using secure quantum channel. Black double lines with white boxes denote classical communications between corresponding parties.}
    \label{fig:2}

\end{figure}

 We generalize the OTT to construct the \textit{quantum one-time table} (QOTT) by replacing the part \textit{(i)} with two maximally entangled states, say, $\ket{\Theta}_{EC}$ and $\ket{\Omega}_{SK}$ and replacing the part \textit{(ii)} with a masking unitary $M_{CS \to AB}$, whose safe state has only one nonzero eigenvalue $1/p$ with degeneracy $p$ for a prime number $p$, and the QOTP cipher, i.e. $X^{j_1i_1}Z^{j_2i_2}$ (X and Z are $p-$dimensional generalized Pauli operators. See Fig. \ref{fig:1}) applied on the system $K$ with numbers $i_1,i_2$ randomly chosen from $\mathds{Z}_p$ and $j_1,j_2$ chosen from some $J \subset \mathds{Z}_p$ . Here,  $(i_1,i_2)$ and $(j_1,j_2)$ should be privately informed respectively to Alice and Bob. (See \textbf{SETUP} of Fig. \ref{fig:2}.) The QOTT could be considered an entangled state partly shared between Alice and Bob through a $(2,3)-$threshold QSS scheme with the system $K$ `locked' with the hidden Pauli operators so that, using the QOTT, any quantum state can be shared as a quantum secret by employing quantum teleportation.

 We propose a qubit-commitment scheme utilizing the QOTT defined above. Although we will introduce a partially credible third party, the \textit{``trusted initializer''} usually named \textit{Ted}, to establish the QOTT in the scheme, we can treat the QOTT as a resource independently of its origin. Introducing a third party does not trivialize our problem, which is the case of the commodity-based cryptography, since the trusted initializer never actively mediates clients, i.e. Alice and Bob, by relaying message transmissions.

\remark{
In our scheme, participants are assumed to be connected to a quantum network, i.e. Alice and Bob, as well as Ted could make use of quantum channels between them.} \remark{In the following description of the scheme, we fix the masking unitary $M_{CS \to AB}$ as the one used in the definition of the QOTT and the entangled state $\ket{\Omega}_{SK}$ as the safe-key state of the quantum masker.} Our scheme consists of three phases \textbf{SETUP}, \textbf{COMMIT} and \textbf{REVEAL}, and Ted participates only in the \textbf{SETUP} phase and exits the scheme afterwards. (See Fig. \ref{fig:2}.)  The protocols are given as follows:

\textbf{SETUP:} Ted prepares a bipartite state $(\mathds{1}_E\otimes M_{CS\to AB} \otimes X_{K}^{j_1i_1}Z_K^{j_2i_2})\ket{\Theta}_{EC}\ket{\Omega}_{SK}$ with randomly chosen indices {$(i_1, i_2)$ from $\mathds{Z}_p \times \mathds{Z}_p$ and $(j_1,j_2)$ from $J\times J$ with some $J \subset \mathds{Z}_p^\times=\{1,\dots,p-1\}$ such that $|J|\geq 2$.} Here, $\ket{\Theta}_{EC}=d^{-\frac{1}{2}}\sum_{i=0}^{d-1}\ket{ii}_{EC}$ is a fixed publicly known $d-$dimensional maximally entangled state. Ted distributes the systems $EA$ to Alice and the systems $BK$ to Bob. Ted privately informs Alice of the indices $(i_1, i_2)$  and {privately} informs Bob of {the indices $(j_1,j_2)$}.

\textbf{COMMIT:} Alice prepares a $d-$dimensional secret state $\rho_I$ in the system $I$. Alice performs a projective measurement onto the basis $\{( X_I^a Z_I^b \otimes \mathds{1}_E)\ket{\Theta}_{IE}\}_{a,b=1}^{d}$ on the systems $IE$ and transmits its outcome $(a,b)$ to Bob.

\remark{
\textbf{REVEAL:} Alice sends the system $A$ to Bob and reveals the indices $i_1$ and $i_2$. Bob applies the gate $Z^{-j_2i_2}X^{-j_1i_1}$ on the system $K$ and then applies the unmasking unitary $M_{AB \to CS}^\dag$ on $AB$. Bob checks if the system $SK$ is in the state $\ket{\Omega}_{SK}$ by implementing a projective measurement $\{\dyad{\Omega}_{SK} , \mathds{1}_{SK}- \dyad{\Omega}_{SK}\}$. If the check is successful, Bob applies $X^a Z^b$ on the system $C$ and accepts the state in the system $C$. If not, Bob rejects.
}

We will call the proposed scheme the \textit{initializer (qubit-commitment) scheme}.

In order to examine the security of a commitment scheme, the \textit{list-approach}, the approach that checks if sensible security criteria such as security against Alice and Bob and their perfectness (see the introduction for definitions) hold, is often taken \cite{Rivest, muller2009composability,kent2012unconditionally,kent1999unconditionally}. However, for qubit-commitment schemes the possibility of committing to a part of entangled state before controlling the rest of it and the impossibility of directly opening an unknown quantum state challenge a satisfying definition of security against Alice.

Therefore, we adopt the stronger \textit{simulation paradigm} \cite{muller2009composability} and set the \textit{delayed quantum teleportation} protocol as our ideal functionality. The delayed quantum teleportation protocol is same as the usual quantum teleportation except that the classical message is fixed in an earlier stage and only its transmission is delayed to the later point of the protocol. Observe that the delayed quantum teleportation achieves the goal of qubit-commitment. (See the \textit{Discussion} section.)

We will say that a qubit-commitment scheme is unconditionally secure if it can be shown, without any assumption on computational power, that the scheme's output is indistinguishable from the output of an instance of the ideal functionality unless the probability of Alice passing the \textbf{REVEAL} phase is lower than a certain threshold that can be made arbitrarily small by increasing security parameters.   We provide the security proof of the schemes in Appendix.

\begin{theorem}
The initializer qubit-commitment scheme is unconditionally secure with the security failure probability upper bounded by $|J|^{-1}.$
\end{theorem}
Essentially, Alice can only try to cheat by reporting wrongful indices $(i_1,i_2)$ to Bob in the \textbf{REVEAL} phase, which leaves $p^2-1$ possible choices for Alice. Among those $p^2-1$ choices, for $2(p-1)$ choices, the success probability of cheating is upper bounded by $|J|^{-1}$ and for $(p-1)^2$ choices, the success probability is upper bounded by $|J|^{-2}$. When Alice reports the correct indices that she received in the \textbf{SETUP} phase, whatever quantum operation she applies to her systems, the already committed state cannot be changed and the probability of Alice being accepted in the \textbf{REVEAL} phase only can drop when Alice deviates from the protocol.

The security of unconditionally secure qubit-commitment scheme is not threatened by the \textit{entanglement attack} that forbids a large class of unconditionally secure (in the list-approach sense) quantum bit-commitment schemes \cite{M,LC}. Two types of attacks based on the properties of entanglement can be considered. The first is committing to a well-defined quantum state and using the fact that the state shared between Alice and Bob is entangled to alter the already committed state. However, already discussed, reporting wrongful indices leads to vanishing success probability, so Alice is forced to report the correct indices $(i_1,i_2)$. In that case, whatever local operation Alice applies on the system $A$, it can only decrease the success probability and cannot alter the committed state a bit, because of the duality of quantum maskers. (See Appendix)

The second type of entanglement attack is committing to a part of entangled state and using the other part for cheating. The security against this type of attack follows from the fact that our security proof relies on the fact that the scheme is indistinguishable (with arbitrarily high probability) from a delayed quantum teleportation, and in quantum teleportation it is allowed to teleport a part of entangled state. In other words, whatever control Alice applies on her part of entangled state after a part of it is committed to, it would only affect her part of quantum state. Note that this can happen even when Alice simply sends her secret state sacrificing the condition that the state should be hidden from Bob until the \textbf{REVEAL} phase begins.

The classical counterpart of committing to a part of entangled state is committing to one of two maximally correlated random variables, for which it is easy to observe that no advantage can be gained from doing that for cheating Bob. See Appendix for more extensive discussion on this type of `pseudo-attack'. It is important that in qubit-commitment Alice does not reveal the classical description of the quantum state she has committed to, since Alice is allowed to commit to an unknown quantum state, like in quantum teleportation.

We also highlight that Ted learns nothing about the secret \remark{quantum state}. This observation ensures that even if Ted is malicious, it is impossible for Ted to attain any information about the secret \remark{quantum state.}

\section{Discussion}

\subsection{Security}
The ideal functionality of our scheme, the delayed quantum teleportation is by definition considered secure. The justification is that in quantum teleportation, we consider the moment when classical communication (the transmission of the outcome of Bell measurement) is done is when the quantum information is transmitted. Therefore, by somehow fixing the classical message to be transmitted and delaying the actual transmission, we can fix the quantum message itself too.

The justification of general forms of Alice's allowed behavior other than the generalized Bell measurement is that even though Alice is demanded to perform the generalized Bell measurement and report the outcome $(a,b)$, to Bob's perspective it is impossible to examine if the generalized Bell measurement was really implemented. Therefore, assuming that an arbitrary subchannel $\Delta_{EA \to A'}^{a,b}$ was applied to Alice's part of the entangled state for each case of reported $(a,b)$ is the best that Bob can do, and the set of $\{\Delta_{EA \to A'}^{a,b}\}$ should be regarded as Alice's legitimate input behavior for quantum teleportation. In this case, Bob is considered to receive the system $C$ of (the normalized version of) the entangled state $ X_C^a Z_C^b \left( \Delta_{EA \to  A'}^{a,b}(\dyad{\Theta}_{EC}\otimes \dyad{\Psi}_{AS})\right) Z_C^{-b}X_C^{-a}$. (Here, appending the ancillary state $\ket{\Psi}_{AS}$ before applying $\Delta_{EA\to A'}^{a,b}$ is also assumed as a specific form of Alice's behavior.) Therefore, applying quantum channel $\Xi_{A' \to A}$ to the system $A'$ and measuring the systems $AS$ are all Alice's local operations happening after the decision of indices $(a,b)$, so they have no causal effect to Bob. In this context, the generalized Bell measurement is only a recommended behavior for Alice to securely commit to her intended quantum state. If she deviates from the protocol in the \textbf{COMMIT} phase, it is simply equivalent for her to deciding to commit to another quantum state.

   Another way of justifying this approach is observing that, even if there is an ideal qubit-commitment protocol (as a black-box), Alice can still commit to a part of an entangled state and apply local measurement on her part of the entangled state and `postselect' by refusing to reveal the committed state whenever she gets an unwanted outcome from the measurement. The possibility of this strategy of Alice is inevitable but it has limited probability of success. Note that between two distrusting parties even a single time of refusal to reveal the secret value could lead to the abortion of communication between them.

`Attacks' of this type are actually not newly introduced by quantum mechanical settings, but are generic in any commitment scheme---including classical bit-commitments. For example, consider the following `attack' on bit-commitment schemes. Suppose Alice commits to a number between 0 and $N-1$ \textit{blindly}, which means Alice does not know which state she has committed to, but keeps a copy of that number in a locked box. In this case one can say that Alice has committed to a part of maximally correlated $N-$dimensional classical bipartite state. Later, Alice `changes' her mind and decides to pretend that she has committed to a particular number, say, 4. Then she opens the locked box hoping that the number is 4---and she succeeds at this `cheating' with probability $1/N$.

One can see that how absurd this `attack' is and the situation is more or less equivalent to the situation in which Alice commits to a part of an entangled state and measures the rest of it at a later point. Post-processing her part of the entangled state ($\Xi_{A'\to A}$ in the proof) would not help her either. This is the reason why we do not count this type of `attacks' (thus, \textit{pseudo-attacks}) as a security failure but consider a legitimate strategy of Alice, and why we consider the delayed quantum teleportation the ideal functionality of qubit-commitment.

From the structure of the security proof, we can see that the initializer scheme is \textit{universally composable}. It is because the security of the scheme is proved in two steps: (i) Substitute a corrupted participant's behavior with an arbitrary quantum operation that outputs the same data as required in the protocol (ii) Show that the whole quantum state is indistinguishable from the output of the ideal functionality, except the cases with negligibly small probability. It fits the definition of security in the universal composability (UC) framework in \cite{muller2009composability,unruh2004simulatable, unruh2010universally,lemus2019minimal}, in the sense that no arbitrary outer machine (called \textit{environment machine} in \cite{muller2009composability}) interacting with the adversary can distinguish the outputs of ideal and real cases. As it was emphasized in \cite{lemus2019minimal}, since qubit- or bit-commitment schemes are used as a basic building block of more complicated cryptographic schemes, it is important to prove its universal composability.

\subsection{Noise and Feasibility}

 \remark{

In experimental aspect, the proposed scheme can be realized with preexisting technologies for quantum teleportation and $(k,n)-$threshold quantum secret sharing \cite{chen2005experimental,bogdanski2008experimental,tittel2001experimental,zhang2005multiparty}, since  Ted in our scheme is simply distributing a part of maximally entangled state as a shared quantum secret using a $(2,3)-$threshold QSS scheme \cite{cleve1999share} and the commitment by Alice is done by quantum teleportation. Noise in realistic situations does not help Alice's dishonest behavior as all the noises are indistinguishable from detectable malicious acts of Alice. Imperfections in Bob's measurement device can be statistically distinguishable from malicious acts of Alice. }

 Any noise happening to the systems possessed by Alice can be considered a part of her behaviors $\Delta_{EA \to A'}$ and $\Xi_{A' \to A}$. Whenever Alice reports the correct indices $(i_1,i_2)$, regardless of the forms of $\Delta_{EA \to A'}$ and $\Xi_{A' \to A}$, the scheme is indistinguishable from the delayed quantum teleportation with noisy apparatus, so the scheme is still secure. When Alice reports wrongful indices, as the upper bound $1/|J|$ of her success probability does not depend on the specific form of the state $\Lambda_{\Delta,\Xi}$ (See Appendix), the scheme is secure nonetheless.
When Bob's measurement device has the probability of $\epsilon$ of having dark count, which means that with probability $\epsilon$ the measurement device clicks without proper input, then it only additively increases the probability of accepting the wrong indices at most by $\epsilon$. If $1/|J| + \epsilon$ can be suppressed to the tolerable failure probability, then the one-shot qubit-commitment is possible. Otherwise, by encoding the secret state using an error correcting code that distributes the state into $n$ systems and implementing the protocol for each system, one can suppress the security failure probability to $(1/|J|+\epsilon)^n$, thus the scheme can be run with additional resources.

\subsection{Shared Randomness Cost}

\remark{Although the QOTT is an entangled state, one cannot simply say that the entanglement in the QOTT provides the unconditional security because quantum communication is already allowed between Alice and Bob. Ted's role is providing uncertainty. Since in a cryptographic setting, mistrustful participants should assume that each other will retain all the side information of their operations, i.e. every information process is isometry, the whole quantum state will remain pure without a publicly trustful source of randomness e.g. intervention of a trusted third party. Therefore we measure the cost of a commodity in terms of its entropy and will call it the \textit{shared randomness cost} (SRC).

One can draw an analogy between shared randomness in the commodity-based cryptography framework and entanglement in the LOCC framework, in the aspect that both are resources in the form of quantum correlation that should be prepared beforehand. The QOTT is, however, different from the conventional resources in quantum information science as mixing of QOTTs only makes it more costly. This does not mean that the more mixed the QOTT is, the more useful it is, since randomness that cannot be verified by distrustful parties is useless in cryptographic setting. Calculating the number of `distillable' QOTTs of an arbitrary bipartite state is an interesting open problem.} 

We first prove an optimality result of Rivest's scheme. Consider a OTT consisting of two correlated random variables $X$ and $Y$ belonging to Alice and Bob respectively. We impose a few conditions for them to be used for commitment schemes.

In a general commodity-based commitment scheme, Alice encodes her $d$-dimensional secret message with the random variable $X$, and sends the encoded message (`commitment') to Bob in the \textbf{COMMIT} phase. In the \textbf{REVEAL} phase, Alice reveals the secret message along with $X$ (`decommitment'). Based on the information $Y$, Bob either accepts or rejects. For this scheme to work properly, $X$ should be random enough to encode a randomly chosen message and, conditioned on $Y$, there should be no ambiguity about the acceptance in the \textbf{REVEAL} phase. Acceptance of $X$ should be based on the conditional probability $Pr(X|Y)$, i.e. Bob, who has $Y=y$, accepts Alice's $x$ only when $x'$ $Pr(X=x'|Y=y)$ is nonzero. For it to be unambiguous, the conditional probability $Pr(X|Y)$ should be uniform on its support. We impose this requirement in the following condition. We let $\Theta$ be the characteristic function on the support of the joint probability $Pr(X=x,Y=y)$, i.e. $\Theta(x,y)=0$ when $Pr(X=x,Y=y)=0$ and $\Theta(x,y)=1$ when $Pr(X=x,Y=y)>0$.

\textbf{Condition (i)} : For every $x$ and $y$, $Pr(X=x|Y=y)=\frac{1}{m}\Theta(x,y)$, where $m\geq d$ is some positive integer.

We remark that $\Theta(x,y)$ functions as an indicator showing if Bob with $Y=y$ accepts Alice's claim that $X=x'$. 

Second, the success probability of cheating by Alice should be bounded from above by some value that vanishes as the security parameter increases. Let $P_{acc}(x'|x)$ be the probability of being accepted by Bob after Alice revealed $x'$ even though she actually received $x$. Note that $P_{acc}(x'|x)=\sum_y \Theta(x',y)Pr(Y=y|X=x)$.

\textbf{Condition (ii)} : For every $x$ and $x'$ such that $x\neq x'$,  $P_{acc}(x'|x)<q$ for some $q=o(1)$.

From these two conditions, we have the following result. See Appendix for proof.

\begin{theorem} \label{thm:optimal}
For an arbitrary OTT (X,Y) satisfying conditions \textit{(i)} and \textit{(ii)}, $Pr(X=x,Y=y)<q^2m^{-1} +o(d^{-2})$ for every $x$ and $y$. 
\end{theorem}
 
 Hence the SRC of (X,Y) is asymptotically lower bounded by $-2\log q + \log d$. For the case of Rivest's scheme, $q=d^{-1}$, and the SRC of the OTT in Rivest's scheme is $3\log d$. Therefore Rivest's scheme achieves the bound of Theorem \ref{thm:optimal}. Even though if the Conditions \textit{(i)} and \textit{(ii)} encompass every OTT capable of implementing commitment scheme is unclear, at least we can see that Rivest's scheme is optimal among a large class of commodity-based commitment schemes.

We compare the SRC of our scheme to that of a qubit-commitment-via-bit-commitment scheme inspired from Ref. \cite{Adcock}, in which Alice sends a quantum state to Bob hidden with the QOTP and commits to the classical OTP part via Rivest's bit-commitment scheme. In this case, the SRC is $6 \log_2 d$ bits for $d-$dimensional qubit-commitment. The success probability of cheating is at most $d^{-1}$ for $2(d-1)$ cases and $d^{-2}$ for $(d-1)^2$ cases out of $d^2-1$ possible cheating strategies.  On the other hand, for the scheme introduced here, the SRC is, at minimum, $2\log_2 d + 2\log_2 |J|$ bits, where $|J|$ is the size of the set of indices used in the scheme. For the given $|J|$, the success probability of cheating by Alice is similarly at most $|J|^{-1}$ for $2(d-1)$ cases and $|J|^{-2}$ for $(d-1)^2$ cases out of $d^2-1$ possible cheating strategies. To achieve the same level of security with Rivest's scheme, we pick the maximal value $|J|=d-1$ and get the asymptotically same level of security with the SRC of $4\log_2 d + o(1)$. This proves that there exists a \textit{quantum} commodity that exhibits a cost advantage for the commodity-based quantum cryptography.

The use of the QOTT is not limited to cryptography of quantum information. The initializer scheme can also be used for boosting efficiency of the commodity-based bit-commitment. Since one can commit to 2 bits of classical information by committing to 1 qubit by employing the superdense coding (first Alice encodes classical information into a maximally entangled state through superdense coding before sending a half of the state to Bob and finally commits to the rest of the state), we have an initializer quantum bit-commitment scheme with SRC of $\log_2d + \log_2|J|$ bits, which is strictly smaller compared to the SRC of Rivest's scheme, $\log_2 d + 2\log_2 |J|$, with the same level of security.

As the SRC reduction by the initializer scheme proposed in this work suggests, quantum strategy can reduce the randomness cost for achieving the same level of security by half. We conjecture that, by improving the scheme with a more clever usage of quantum advantage, one can reduce the randomness cost of hiding information too, hence the overall SRC can achieve (asymptotically) $3\log_2 d$ bits for the commitment of $d$-dimensional quantum state.

Our qubit-commitment scheme, satisfying general security criteria while preserving coherence of the committed quantum state, has broad applications in the upcoming era of quantum network \cite{kimble2008quantum}. One example could be an implementation of a fair version of \textit{quantum state exchange} \cite{lee2019state,oppenheim2005uncommon}. If two quantum computers are demanded to generate certain quantum states independently and to be cross-checked afterward \cite{elben2020cross}, one computer should not acquire the other's output state before generating its state. Otherwise, one could learn from the other's state or even try to imperfectly clone \cite{buvzek1996quantum} the state to pretend to have high computational power. Our qubit-commitment can solve this problem by letting the first revealer persuade the other party that it indeed has generated the quantum state independently without giving the information of the state before receiving the other party's quantum state. In general, qubit-commitment can lift the assumption of \textit{asynchronous quantum network} \cite{unruh2004simulatable}, i.e. the assumption that only one quantum machine among many connected to a quantum network can be activated at a time, for quantum cryptographical models, since effectively multiple quantum machines can be activated at the same time utilizing qubit-commitment.

\begin{acknowledgments} 
SHL is grateful to P. Boes, F. Buscemi, C. Oh, S. Shin, S. Choi and K.C. Tan for insightful discussions. This work was supported by National Research Foundation of Korea grants funded by the Korea government (Grants No. 2019M3E4A1080074, No. 2020R1A2C1008609 and No. 2020K2A9A1A06102946) via the Institute of Applied Physics at Seoul National University and by the Ministry of Science and ICT, Korea, under the ITRC (Information Technology Research Center) support program (IITP-2020-0-01606) supervised by the IITP (Institute of Information \& Communications Technology Planning \& Evaluation).
HK and MSK are supported by the Korea Institute of Science and Technology Institutional Program (2E26680-18-P025), a Samsung Global Research Outreach project, the Royal Society and the QuantERA ERA-NET Cofund in Quantum Technologies implemented within the European Union's Horizon 2020 Programme.
\end{acknowledgments} 

\onecolumn \newpage

\appendix
\section{PROOF OF THE RESULTS}
\newtheorem{appth}{Theorem}
 \begin{appth}
 Let $\omega_S$ be the safe state of a universal quantum masker $\Phi_M$ for d-dimensional quantum system. Then, the von Neumann entropy of $\omega_S$ is lower bounded by $\log_2d$.
 \end{appth}
 
 Before proving this theorem we prove a more general result for $(k,n)-$threshold quantum secret sharing schemes given in \cite{imai2003quantum} in a way that doesn't rely on the strong subadditivity of the von Neumann entropy.
 \begin{lem}
 Assume that $\Phi_M:\mathcal{B}(\mathds{C}^d) \to \mathcal{B}((\mathds{C}^d)^{\otimes n})$ is the quantum map implementing a $(k,n)-$threshold quantum secret sharing scheme. Then, for any $\rho \in \mathcal{B}(\mathds{C}^d)$, the marginal state $\Phi_M(\rho)_{A_i}$ of any system $A_i$ obtained by tracing out the other $n-1$ parties of the $n$-partite state $\Phi_M(\rho)_{A_1,\dots ,A_n}$ has the von Neumann entropy of $\log_2 d$ or higher.
 \end{lem}
\begin{proof}
As every $(k,n)-$threshold quantum secret sharing scheme can be purified to a pure $(k,2k-1)$-threshold quantum secret sharing scheme \cite{imai2003quantum}, we only prove the lemma for that case. In that case, the scheme can be implemented with an isometry $M:\mathds{C}^d \to (\mathds{C}^d)^{\otimes 2k-1}$. Consider the input state $\rho \in \mathcal{S}(\mathds{C}^d)$ in the system $C$ and its purification $\ket{\Psi_\rho}_{EC}$ with the purification system $E$. Then the isometry $M$ outputs the state $(\mathds{1}\otimes M)\ket{\Psi_\rho}_{EC}$ distributed among $2k$ parties $E$ and $A_1,...A_{2k-1}$. Let $D$ be an authorized subset of the parties $\{A_1,\dots,A_{2k-1}\}$ of the size $k$ containing $A_i$ and $U$ be the unauthorized subset $\{A_1,\dots,A_{2k-1}\} \setminus D$ of the size $k-1$. As a secret sharing scheme decouples any unauthorized set from environment, we have
    $$H(E,U)=H(E)+H(U),$$
    where all the von Neumann entropies are defined for the output state $(\mathds{1}\otimes M)\ket{\Psi_\rho}_{EC}$. Since $(\mathds{1}\otimes M)\ket{\Psi_\rho}_{EC}$ is a pure state we have
    $$H(E,U)=H(D),$$
    and from the subadditivity of the von Neumann entropy we have
    $$H(D)\leq H(A_i) + H(U')$$
    where $U':= D \setminus \{A_i\}$ is an unauthorized subset of $\{A_1,\dots,A_{2k-1}\}$ of the size $k-1$. Therefore,
    $$H(E)+H(U)-H(U')\leq H(A_i).$$
    Since the choice of the authorized set $D$ was arbitrary barring the condition that it contains $A_i$, one can choose the new $D$ as $\{A_i\}\bigcup U$, which makes the new $U'$, $U$. From the same argument we have
    $$H(E)+H(U')-H(U)\leq H(A_i).$$
    By averaging two inequalities we have
    $$H(E)\leq H(A_i).$$
    As the system $E$ was defined as the purifying system of the input state $\rho$ and remained intact through the secret sharing process, $H(E)=H(\rho).$ As $\rho$ was arbitrarily chosen, however, we can take $H(E)$ as its maximum value $\log_2 d$.

\end{proof}
Note that this result can be considered a stronger version of Theorem 4 in \cite{imai2003quantum}. By noting that a quantum masking process is merely a $(2,2)-$threshold quantum secret sharing scheme and the fact that any mixed quantum secret sharing scheme can be obtained by tracing out irrelevant parties of a pure quantum secret sharing state, we can see that the following circuit represents an implementation of a pure $(2,3)-$threshold quantum secret sharing scheme among three parties, $A,B$ and $K$.
 \begin{equation}\nonumber
         \Qcircuit @C=.8em @R=.9em{
    \lstick{}&\qw&\qw&\rstick{E}\qw \\
    \lstick{}&\multigate{1}{M_{CS \to AB}}&\qw&\rstick{A}\qw\\
    \lstick{}&\ghost{M_{CS \to AB}}&\qw&\rstick{B}\qw\\
    \lstick{}&\qw&\qw&\rstick{K}\qw
    \inputgroupv{1}{2}{.8em}{.8em}{\ket{\Psi_\rho}_{EC} \;\;\;\;\;}
    \inputgroupv{3}{4}{.8em}{.8em}{\ket{\Omega}_{SK} \;\;\;}
    }
 \end{equation}
 Note that $\ket{\Psi_\rho}_{EC}$ is a purification of the input state $\rho$ and $\ket{\Omega}_{SK}$ is a purification of the safe state $\omega_S$ consumed in the hiding process. The lemma says that $H(K)\geq \log_2 d$, but as $H(K)=H(\omega_S)$ we have the proof of the theorem 1.
\begin{appth}
Let $\omega_S$ be the safe state of a universal quantum masker $\Phi_M$ for d-dimensional quantum system. If the masked state $\Phi_M(\rho)$ for any $\rho$ is separable, then the von Neumann entropy of $\omega_S$ should be strictly larger than $\log_2 d$.  If the quantum discord $D^{\leftarrow}(A|B)$ of the masked state $\Phi_M(\rho)$ in one direction for any state $\rho$ vanishes, then the von Neumann entropy of $\omega_S$ is at least $2 \log_2 d$.
\end{appth}
\begin{proof}
Let's say $\Phi_M(\rho)$ is separable for any pure state input $\rho$, then the fact following equality holds is evident from the definition of quantum discord.
$$S(K)=I(A:B)+D^{\leftarrow}(A|K)+D^{\leftarrow}(B|K).$$
Here $I(A:B)=S(A)+S(B)-S(AB)$ is the quantum mutual information betwenn $A$ and $B$. However, as at least one of the pairs $AK$ and $BK$ should be entangled for some pure state $\rho$, because if all three pairs $AB$, $AK$ and $BK$ are separable, then the bipartite staet $\Phi_M(\rho)_{AB}$ should be classically correlated \cite{thapliyal1999multipartite}, but because of Theorem 4 below, it is impossible to mask quantum information into classically correlated systems. Note that quantum discord $D^{\leftarrow}(A|K)$ is zero if and only if the system $AK$ is in a quantum-classical state. Therefore for the given conditions, $S(K)$ must be strictly larger than $I(A:B)$, which is in turn no smaller than $\log_2 d$. As $S(S)=S(K)$, we get the first part of the theorem.

For the second part of the theorem, as $D^{\leftarrow}(A|B)=0$, the state $\Phi_M(\dyad{\psi})$ is a quantum-classical (QC) state \cite{datta2010condition} that has the form, if $\Tr_A (\Phi_M(\dyad{\psi}))=\sum_i q  _i \dyad{\sigma_i}_B$ is the spectral decomposition of the state of the system B independent of the state $\dyad{\psi}$,
$$\Phi_M(\dyad{\psi})=\sum_i q_i \mathcal{M}_i(\dyad{\psi})_A\otimes\dyad{\sigma_i}_B,$$
where  $\mathcal{M}_i$ is some quantum process for each $i$. Let's say $\rho$ and $\sigma$ are arbitrary quantum states that have orthogonal support ($\rho \sigma = 0$). Then we have 
$\Tr [{\Phi_M(\rho)}{\Phi_M(\sigma)}] = \sum_i q_i^2 \Tr[\mathcal{M}_i(\rho)\mathcal{M}_i(\sigma)] = \Tr[M (\rho_C \otimes {\omega_S})M M^\dag (\sigma_C \otimes {\omega_S})M^\dag]=\Tr[\omega_S^2]\Tr[\rho \sigma]=0$. Therefore $\Tr[\mathcal{M}_i(\rho)\mathcal{M}_i(\sigma)]=0$ for all $i$ such that $q_i\neq 0$. Therefore for any Hermitian operator $H$, we can decompose it into the positive part $P\geq 0$ and the negative part $N \geq 0$ that are mutually orthogonal so that $H=P-N$. This leads to $\|\mathcal{M}_i(H)\|_1=\|\mathcal{M}_i(P)-\mathcal{M}_i(N)\|_1=\|\mathcal{M}_i(P)\|_1+\|\mathcal{M}_i(N)\|_1=\|P\|_1+\|N\|_1=\|P-N\|_1=\|H\|_1,$ where we used the fact $\|\mathcal{M}_i(\rho)\|_1=\|\rho\|_1$ for any positive operator $\rho$ since $\mathcal{M}_i$ is a CPTP (Completely positive and trace preserving)  map. This proves that $\mathcal{M}_i$ preserves the trace norm on the space of $d$-dimensional Hermitian operators for all $i$ with $q_i \neq 0$ so that it is injective. It follows that all $\mathcal{M}_i$ are invertible quantum maps. From the masking condition, the quantum channel $\Tr_B[\Phi_M(\;\cdot\;)]=\sum_i q_i\mathcal{M}_i(\;\cdot\;)$ is a \textit{randomization scheme} \cite{NS} and therefore, from the result of the Ref. \cite{NS}, the Shannon entropy of the probability distribution $\{q_i\}$, which is smaller than the von Neumann entropy of the safe state $\omega_S$ as $\Phi_M(\dyad{\psi})=M_{CS\to AB} (\dyad{\psi}_C \otimes \omega_S)M^\dag_{AB \to CS}$, is at least $2\log_2 d$.
\end{proof}

Indeed the randomness lower bounds proved above are indeed minimums by the existence of 4-qubit masker and quantum one-time pad.

\begin{lem}
  For arbitrary $d\geq2$, universal quantum masker exists for a $d$-dimensional quantum system that consumes $\log_2d$ bits of randomness.
\end{lem}
\begin{proof}

The 4-qubit masker can be easily generalized to $d$-dimensional systems by replacing the 2-dimensional controlled-X gates with its $d$-dimensional generalization given as
$$U_{C-X}\ket{x}\ket{y}=\ket{x}\ket{x\oplus y \;(\text{mod}\; d)},$$
for $x,y=0,...,d-1$ and replacing the Hadamard gate with the discrete Fourier transform gate,
$$U_{DFT}=\sum_{i=0}^{d-1}\dyad{\tilde{i}}{i},$$
where$\ket{\tilde{n}}:=\sum_{j=0}^{d-1}\exp(i2\pi nj/d)\ket{j},$
for $n=0,...,d-1$. In this case, the output state for a pure input state
$\ket{\psi}=\sum_{i=0}^{d-1}\alpha_i\ket{i},$ is given as
$$\rho_{output}=\frac{1}{d}\sum_{i=0}^{d-1} \dyad{\Psi_i}{\Psi_i}_{A_1B_1}\otimes\dyad{\Phi_i}{\Phi_i}_{A_2B_2},$$
where
$$\ket{\Psi_n}:=\sum_{j=0}^{d-1} \alpha_{n\oplus j\;(\text{mod}\;d)}\ket{j}\otimes\ket{j},$$
and
$$\ket{\Phi_n}:=\frac{1}{\sqrt{d}}\sum_{j=0}^{d-1} e^{i\frac{2\pi jn}{d}}\ket{j}\otimes\ket{j},$$
for $n=0,...,d-1$.

For every $d\geq 2$, tracing out $B_1 B_2$ system yields the maximally mixed state for the system $A_1 A_2$ and \textit{vice versa}. For general mixed states, from the linearity of quantum processes, the marginal state of the output state will be a mixture of maximally mixed marginal states, which is again the maximally mixed state. This shows that the given quantum process can mask any quantum state. As to the recoverability condition, since this operation consists of unitary operation after the attachment of ancillary system, simple inverse unitary operation followed by tracing out of the ancillary systems recovers the input system.
\end{proof}

\begin{lem}
Universal quantum masking is impossible without quantum correlation.
\end{lem}
\begin{proof}
For a universal masking process $\Phi_M$, having no quantum correlation in the masked quantum state implies the following expression for any input state $\rho_C$.
$$\Phi_M(\rho) = \sum_{ij}p_{ij}(\rho) \dyad{i}_A\otimes\dyad{j}_B,$$
where $p_{ij}(\rho)$ is a joint probability distribution for indices $i$ and $j$ linear in the state $\rho$ and $\{\ket{i}_A\}$ and $\{\ket{j}_B\}$ are respectively eigenbasis of $\Tr_B[\Phi_M(\rho)]$ and $\Tr_A[\Phi_M(\rho)]$ which are independent of the input state $\rho$. From the proof of Theorem 2, however, $\Phi_M(\rho)$ also permits the following expression for any input state $\rho$.
$$\Phi_M(\rho)=\sum_i q_i \mathcal{M}_i(\rho)\otimes \dyad{i}_B.$$
By letting $\mathcal{N}_i(\rho):=(1/q_i)\sum_j p_{ji}(\rho)_A\dyad{j}_A$ for every $i$, we have $\Im{\mathcal{N}_i }= \Im{\mathcal{M}_i}$. The left hand side, however, is diagonal in the basis $\{\ket{i}_A\}$ but the right hand side is an isometric embedding of the space of quantum states $\{\rho_C\}$, which is a contradiction.
\end{proof}
\begin{appth}
The initializer qubit-commitment scheme is unconditionally secure.
\end{appth}

\begin{proof}
Although we have adopted the simulation paradigm, we can still check if two important security criteria for commitment schemes are satisfied. \textit{Correctness} condition, i.e. that when all participants are honest, the outcome state revealed to Bob is the same with the secret state that Alice committed to, is satisfied trivially since $M_{CS \to AB}$ and $M_{CS \to AB}^\dag$ cancel each other. \textit{Security against Bob} holds since, without knowledge of indices $(i_1,i_2)$, the state ${(\mathds{1}_S\otimes X^{j_1i_1}Z^{j_2i_2})}\ket{\Omega}_{SK}$ is twirled to $\omega_S \otimes \omega_K$ since $i \mapsto ij$ modulo $p$ is a permutation unless $j \equiv 0 $ (mod $p$). Then, the systems in Bob's possession after the \textbf{COMMIT} phase are in the state (assuming that there is a well-defined secret state) $\Tr_A[M(\rho_C \otimes \omega_S)M^\dag] \otimes \omega_K$, where $\rho_C$ is the state that Alice committed to. From the masking property of $M$, this state is independent of $\rho_C$, therefore the scheme is unconditionally and perfectly concealing. We can also show that the scheme \textit{secure against Alice}  when there is a well-defined secret state in the \textbf{COMMIT} phase. Assume that Alice commits to a pure quantum state $\ket{\psi}$ that is in a product state with its environment. However, at a later point in time, suppose that Alice decides to change the already committed state so she applies a quantum channel $\Xi_{A}$ on the system $A$. But, assuming that Alice reports the correct indices in the \textbf{REVEAL} phase, (otherwise she has arbitrarily small success probability) the output result is always $\ket{\psi}$ and the probability of acceptance in the \textbf{REVEAL} is $\bra{\Psi}_{AS}(\Xi_A \otimes \mathcal{I}_S)(\dyad{\Psi}_{AS})\ket{\Psi}_{AS}$. Therefore, deviating from the protocol after honestly committing to a certain quantum state only leads to decrease of the success probability and cannot change the committed state even slightly.

Next, we claim that whenever Alice reports the rightful indices $(i_1,i_2)$ which are received from Ted, the unnormalized outcome state in the system $C$ is indistinguishable from an outcome of an instance of the \textit{delayed quantum teleportation} scheme in which the classical information transmission is faithful but delayed. This means that the classical information (the outcome of the Bell measurement) is irreversibly decided in the initial stage of the scheme but its revelation is delayed to a later point of time. In the delayed quantum teleportation, Alice can teleport a half of a maximally entangled state and measure the other half to collapse the already teleported state between two stages, but this is not necessarily a malicious act but only a manifestation of the nonlocal property of entangled state. The measurement outcome on Alice's side in the delayed quantum teleportation is probabilistic and cannot be deterministically controlled. If it were possible to control it, then Alice can effectively collapse Bob's system into a state beyond the statistics according to Bob's marginal state (of $\ket{\Theta}_{EC}$) without additional transmissions, thus it leads to the violation of the no-signaling theorem. Therefore, we consider the delayed quantum teleportation as our ideal functionality. (See Discussion section of the paper.)

\begin{figure}
    
    \includegraphics[width=.3\textwidth]{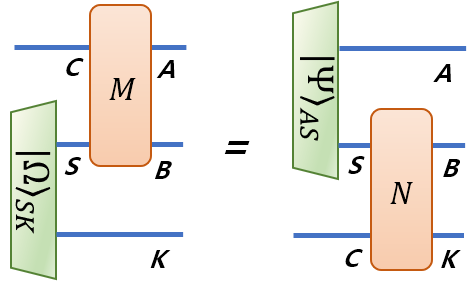}
    \centering
    \caption{Duality of masking unitary operator. If $M_{CS \to AB}$ is the masking unitary of a quantum masker and $\ket{\Omega}_{SK}$ is the corresponding, then the purification of the quantum masker, which is a $(2,3)-$ threshold quantum secret sharing (QSS) protocol, $(M_{CS \to AB} \otimes \mathds{1}_K )(\mathds{1}_C\otimes\ket{\Omega}_{SK})$  , can be equivalently expressed with a masking unitary $N_{CS\to BK}$ and the corresponding safe-key state $\ket{\Psi}_{AS}$ as $(\mathds{1}_A \otimes N_{CS \to BK})(\mathds{1}_C\otimes\ket{\Psi}_{AS})$.  }
    \label{fig:3}

\end{figure}

In our scheme, once the correct indices $(i_1,i_2)$ are reported to Bob in the \textbf{REVEAL} phase, the generalized Pauli gates on the system $K$ are exactly cancelled out in the \textbf{REVEAL} phase so we can ignore them throughout the whole protocol. Although Alice is demanded to perform the generalized Bell measurement in the \textbf{COMMIT} phase, since Bob cannot examine the behavior of Alice other than the measurement outcomes $(a,b)$, the most general behavior of Alice is applying arbitrary subchannels $\Delta_{a,b} : EA \to A'$ such that $\Delta_{EA \to A'} = \sum_{a,b} \Delta_{EA\to A'}^{a,b}$ is a quantum channel (CPTP map) and reporting corresponding indices $(a,b)$ to Bob. Here the system $A'$ can be chosen arbitrarily as it will never be revealed to Bob. In the \textbf{CHEAT} phase, the unofficial phase between the \textbf{COMMIT} and the \textbf{REVEAL} phase, Alice can apply an arbitrary quantum channel $\Xi_{A'\to A}$ of her choice and give the system $A$ to Bob in the \textbf{REVEAL} phase. The unnormalized post-measurement state of the system $C$ when the test in the \textbf{REVEAL} phase is passed is

$$\sum_{a,b=1}^d X^a Z^b \bra{\Omega}_{SK}M^\dag (\Xi_{A' \to A} \circ \Delta_{EA \to  A'}^{a,b})(\dyad{\Theta}_{EC}\otimes M\dyad{\Omega}_{SK}M^\dag)M\ket{\Omega}_{SK} Z^{-b}X^{-a} \otimes \dyad{a,b}_{T}. $$

Here, $M$ is $M_{CS\to AB}$ and $\dyad{a,b}_T$ denotes the classical information Bob received in the \textbf{COMMIT} phase.

However, the masking unitary $M_{CS\to AB}$ has duality because of the no-hiding theorem (See Fig. \ref{fig:3}.) It means that there exists a masking unitary $N_{CS\to BK}$ and the corresponding safe-key state such that $(M_{CS \to AB} \otimes \mathds{1}_K )(\mathds{1}_C\otimes\ket{\Omega}_{SK}) = (\mathds{1}_A \otimes N_{CS \to BK})(\mathds{1}_C\otimes\ket{\Psi}_{AS})$). This follows from the no-hiding theorem since, as the quantum masker hides the information from the system $A$, it should be encoded in its purifying system $BK$ with some invertible quantum map. From the same logic applied to the derivation of the existence of the masking unitary $M_{CS\to AB}$ and the safe-key state $\ket{\Omega}_{SK}$, we can derive the existence of the masking unitary $N_{CS \to BK}$ and the corresponding safe-key state $\ket{\Psi}_{AS}$. Thus, the unnormalized post-measurement state after the \textbf{REVEAL} phase can be rewritten as

$$\sum_{a,b=1}^d X^a Z^b \bra{\Psi}_{AS}(\Xi_{A' \to A} \circ \Delta_{EA \to  A'}^{a,b})(\dyad{\Theta}_{EC}\otimes \dyad{\Psi}_{AS})\ket{\Psi}_{AS} Z^{-b}X^{-a} \otimes \dyad{a,b}_{T}. $$
Note that the masking unitary $N_{CS \to BK}$ does not appear in this expression because it was cancelled out.

However, this unnormalized state can be equivalently considered an unnormalized outcome state of an instance of quantum teleportation in which $\ket{\Theta}_{EC}$ is used as an entanglement resource and the $\ket{\Psi}_{AS}$ is an ancilla of local operations of Alice. (Therefore, in this delayed quantum teleportation scenario, Bob only possesses the system $C$.) The state can be considered the unnormalized post-measurement state of the system $C$ after the ancillary systems $AS$ are measured to stay in $\ket{\Psi}_{AS}$. In this sense, we can say that our scheme is secure whenever Alice reports the correct indices $(i_1,i_2)$. 

It remains to show that reporting wrongful indices has arbitrarily small success probability of passing the check in the \textbf{REVEAL} phase for Alice. For simplicity, we define $\ket{\Omega_{ab}}_{SK}:=(\mathds{1}_S \otimes X^aZ^b)\ket{\Omega}_{SK}$. Note that $\{\ket{\Omega_{ab}}\}_{a,b=1}^p$ forms an orthonormal basis of $\mathds{C}^p \otimes \mathds{C}^p$. We claim that the probability of Alice passing the \textbf{REVEAL} phase, i.e. dual unmasking unitary $N_{BK \to CS}^\dag$ is applied and the system $AS$ is checked if it is in the state $\ket{\Psi}_{AS}$, is same as the probability of the check in which $N_{BK \to CS}^\dag$ is replaced with $M_{AB \to CS}^\dag$ and $\ket{\Psi}_{AS}$ is replaced with $\ket{\Omega}_{SK}$ since $M_{CS \to AB}\ket{\Omega}_{SK}= N_{CS \to AK} \ket{\Psi}_{AS}$ holds from their duality. From the protocols of the scheme, when Alice reports indices $(k_1,k_2)$ instead of $(i_1,i_2)$, the success probability becomes $\bra{\Omega_{j_1(i_1-k_1),j_2(i_2-k_2)}}\Lambda_{\Delta, \Xi}\ket{\Omega_{j_1(i_1-k_1),j_2(i_2-k_2)}}$ for any given $(j_1,j_2)$, where $\Lambda_{\Delta, \Xi}$ is defined as $\Lambda_{\Delta, \Xi} = \Tr_C \left[(\Xi_{A' \to A} \circ \Delta_{EA \to  A'})(M \dyad{\Theta}_{EC}\otimes \dyad{\Omega}_{SK} M^\dag)\right]$. Here, $\Delta_{EA \to A'} = \sum_{a,b} \Delta_{EA\to A'}^{a,b}$ and $\Xi_{A' \to A}$ are arbitrary behaviors of Alice as it was mentioned before. Since $(j_1,j_2)$ is picked from $J \times J$ uniformly, the total security failure probability is

$$\frac{1}{|J|^2}\sum_{j_1,j_2 \in J} \bra{\Omega_{j_1(i_1-k_1),j_2(i_2-k_2)}}\Lambda_{\Delta, \Xi}\ket{\Omega_{j_1(i_1-k_1),j_2(i_2-k_2)}}\leq \frac{1}{|J|} \Tr \left[ \Lambda_{\Delta, \Xi} \right]=\frac{1}{|J|},$$

under the assumption only one of $i_1\neq k_1$ or $i_2 \neq k_2$ holds (for which there are $2(p-1)$ possible cases). Note that if both of $i_1$ and $i_2$ are wrongfully reported (for which there are $(p-1)^2$ possible cases), then the bound becomes $\frac{1}{|J|^2}$. We also remark that the specific form of $\Lambda_{\Delta, \Xi}$ was irrelevant for this derivation and only that it is a quantum state was used. By increasing $|J|$, (necessarily also by increasing $p$,) one gets arbitrarily low security failure probability.

We remark here that although the security proof given here does not explicitly use the notion of the min- or max- entropy or the epsilon-delta argument, but since the proof holds for singe-shot implementation of the protocol and that the maximum success probability of cheating is bounded from above by a parameter that can be lowered arbitrarily by employing more resources, we argue that the proof still holds. Note that the proof given by Rivest \cite{Rivest} also does not use the notions of the min- or max-entropy as well.

\end{proof}

\begin{appth}
For an arbitrary OTT (X,Y) satisfying conditions \textit{(i)} and \textit{(ii)}, $Pr(X=x,Y=y)<q^2m^{-1} +o(d^{-2})$ for every $x$ and $y$. 
\end{appth}
\begin{proof}
    We first find an upper bound of $Pr(X)$. From Condition \textit{(ii)} we get that $\sum_y Pr(X=x'|Y=y)Pr(Y=y|X=x) < \frac{q}{m},$ for any $x \neq x'$. Note that $\sum_y \Theta(x,y)Pr(Y=y|X=x)=1$. By multiplying by $\Pr(X=x)$ and summing over $x$ both sides, we get
    $$ \sum_y Pr(X=x'|Y=y)Pr(Y=y) = Pr(X=x') < (1-Pr(X=x'))\frac{q}{m} + Pr(X=x')\frac{1}{m}. $$
    It implies that $Pr(X=x') < \frac{q}{m-1+q}$ for any $x'$, thus we found an upper bound of $Pr(X)$.
    
    Next, we find an upper bound of $Pr(Y)$. We use another expression of Condition \textit{(ii)}, that
    $$\sum_y \Theta(x',y)\Theta(x,y)Pr(Y=y) < qm Pr(X=x),$$
    for any $x \neq x'$, which follows from $Pr(Y=y|X=x)=Pr(X=x|Y=y)Pr(Y=y)/Pr(X=x)=\frac{1}{m}\Theta(x,y)Pr(Y=y)/Pr(X=x)$. Using the upper bound $Pr(X=x)<\frac{q}{m-1+q}$, we get, for any distinct $x$ and $x'$,
    $$\sum_y \Theta(x',y)\Theta(x,y)Pr(Y=y) < q^2\frac{m}{m-1+q}.$$
    Condition \textit{(i)} implies that there are $m$ $x$'s such that $\Theta(x,y)=1$ for a given $y$. Therefore, for arbitrary $y_0$ we can find two $x$ and $x'$ such that $\Theta(x,y_0)=\Theta(x',y_0)=1$. Since the sum of nonnegative terms is always not smaller than individual terms, we get
    $$Pr(Y=y_0)<q^2\frac{m}{m-1+q}.$$ As $y_9$ was arbitrarily chosen, we get that $Pr(Y)<q^2\frac{m}{m-1+q}$.
    
    Finally, since $Pr(X=x,Y=y)=Pr(X=x|Y=y)Pr(Y=y)\leq\frac{1}{m}Pr(Y=y),$ we have
    $$Pr(X=x,Y=y)<q^2m^{-1}\frac{m}{m-1+q}.$$
    Since $\frac{m}{m-1+q}=1+O(m^{-1})$, $m=O(d)$ and $q=o(1)$, we have $Pr(X=x,Y=y)<q^2m + o(d^{-2})$.
\end{proof}

\bibliographystyle{unsrtnat}
\bibliography{main}
\end{document}